\documentclass[useAMS]{mn2e}
\usepackage{epsfig}
\title[A new LMC K-band distance from precision measurements of nearby
red clump stars]{A new LMC K-band distance from precision measurements
of nearby red clump stars}
\author[C. D. Laney, M. D. Joner and G. Pietrzy\'{n}ski]{C. D.
Laney$^{1}$,$^{2}$\thanks{E-mail:
cephradius@yahoo.com (CDL); jonerm@forty-two.byu.edu (MDJ) and
pietrzyn@sirius.astrouw.pl (GP)}, M. D. Joner$^{1}$ and G.
Pietrzy\'{n}ski$^{3}$,$^{4}$\\
$^{1}$Dept. of Physics and Astronomy, N283 ESC, Brigham Young University, Provo, UT
84601, USA\\
$^{2}$South African Astronomical Observatory, P.O. Box 9, Observatory 7935,
South Africa\\
$^{3}$Universidad de Concepci\'{o}n, Departamento de Astronomia, Casilla 160-C,
Concepci\'{o}n, Chile\\
$^{4}$Warsaw University Observatory, Al. Ujazdowskie 4, 00-478 Warsaw,
Poland}
\begin{document}

\date{Revised text accepted September 12, 2011}

\pagerange{\pageref{firstpage}--\pageref{lastpage}} \pubyear{2011}

\maketitle

\label{firstpage}

\begin{abstract}
High-precision ($\sigma_{mag}<$0.01) new JHK observations of 226 of the
brightest and nearest red clump stars in the solar neighbourhood are used
to determine distance moduli for the LMC. The resulting K- and H-band values
of 18.47$\pm$0.02 and 18.49$\pm$0.06 imply that any correction to the K-band
Cepheid PL relation due to metallicity differences between Cepheids in the LMC
and in the solar neighborhood must be quite small.
\end{abstract}

\begin{keywords}
distance scale -- Magellanic Clouds -- infrared: stars -- stars: variables:
Cepheids
\end{keywords}

\section{Introduction}

In principle, the helium-burning red clump stars as defined by
Paczy\'{n}ski \& Stanek (1998) offer real advantages as distance
indicators. They are a relatively numerous, well-defined population, and
hundreds of red clump stars with quite accurate parallaxes can be found in
the Hipparcos catalog. At first, attention centred on distance determination
using the I band (Paczy\'{n}ski \& Stanek 1998, Stanek \& Garnavich 1998,
Udalski et al. 1998, Udalski 2000), but the effects of stellar population
differences on the mean V-band or I-band red clump magnitude can be
considerable (Alves et al. 2002, Grocholski \& Sarajedini 2002,
Girardi \& Salaris 2001, Groenewegen 2008, Pietrzy\'{n}ski et al. 2010).
In the K band, the effects of stellar population differences and reddening are
generally less (Salaris \& Girardi 2002, Alves et al. 2002,
Grocholski \& Sarajedini 2002, Pietrzynski et al. 2010), although not always
negligible. In particular, the estimated corrections are predicted to be
much smaller when comparing the red clump populations in the solar 
neighbourhood and in the LMC field (Salaris \& Girardi 2002), and the data 
suggest that this is indeed the case 
(Alves et al. 2002, Pietrzy\'{n}ski et al. 2010).

But ever since Alves (2000) first determined a mean K-band absolute magnitude
for nearby red clump stars, a fundamental weakness of this approach has been
the quality of the infrared photometry available for nearby red clump stars.
As pointed out by Alves, the stars with the best Hipparcos parallaxes are all
saturated in the 2MASS survey data, and modern IR array detectors are too
sensitive for stars with K$<$5. Indeed, a typical IR telescope/array
combination like the InfraRed Survey Facility (IRSF) in South Africa has a
bright limit of K$=$8 despite a telescope aperture of only 1.4m. The catalog
data used by Alves were therefore a miscellaneous collection on no
well-defined system, and the more modern data used by Groenewegen (2008)
(giving a rather different result) were restricted to fainter stars.

Here it may be useful to quote Groenewegen:
{\it To settle the issue on the importance of the bias and the absolute
K-magnitude of RC stars would require accurate NIR magnitudes of a 100
to a few hundred (cf. Table 2) bright (K $\sim$ 5)
RC stars. Given the brightness, this represents a challenge to
modern instrumentation because of saturation.}

For this study we have determined accurate K-band magnitudes for 226 bright,
nearby red clump stars with magnitudes brighter than K $\sim$ 5. With these
data we have determined the mean K-band absolute magnitude for red clump
stars in the solar neighbourhood to within 2\%.

\section{Observations and Errors}

\begin{figure*}
  \epsfig{file=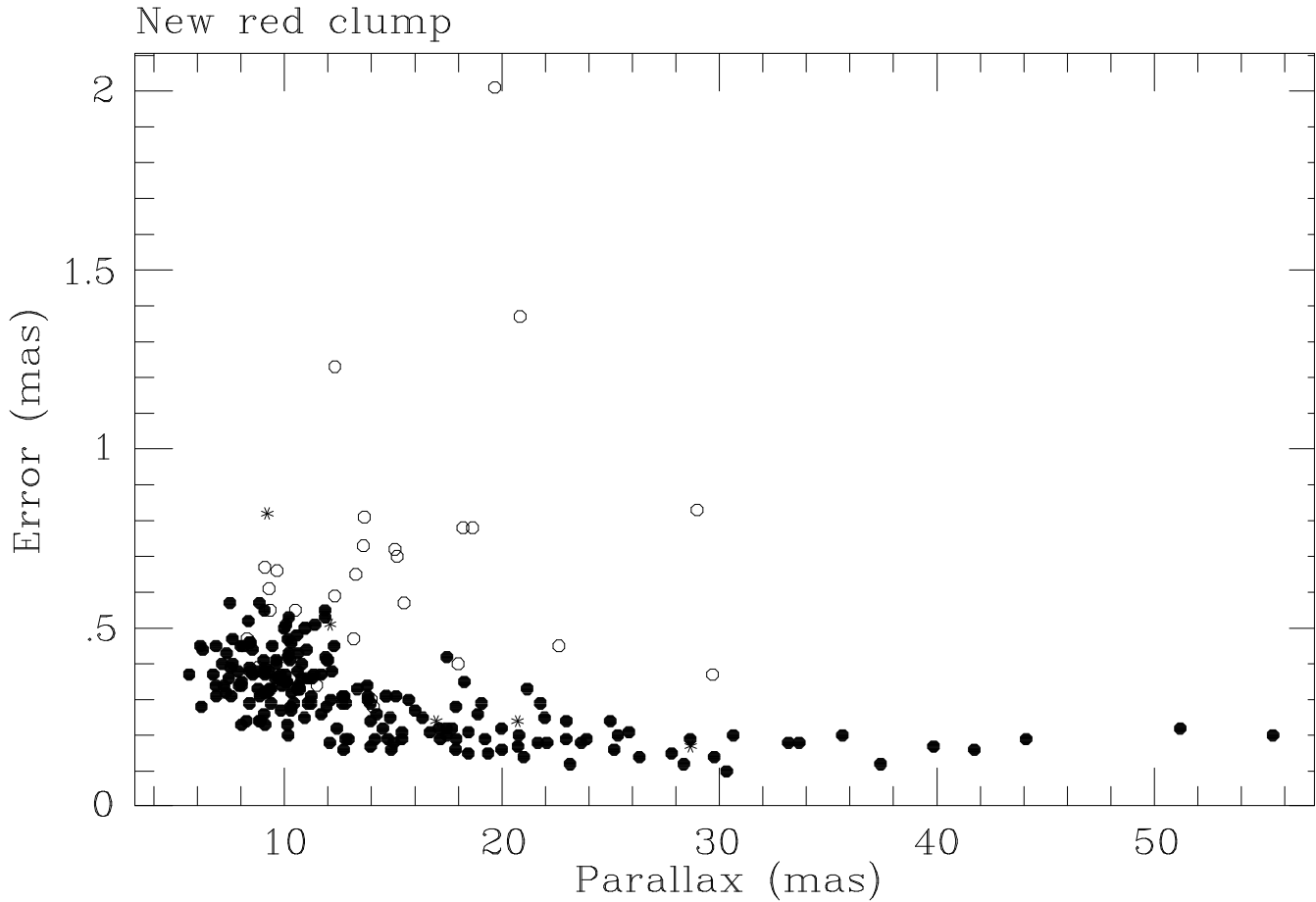, width=\linewidth}
  \caption{Hipparcos parallax vs. error, both in milliarcseconds (mas).
Stars with 5-parameter parallax
fits which were  included in the 'good'' sample  (see text) are represented
by filled circles. Stars with 5-parameter parallax fits which were not
included are represented by asterisks, and stars without 5-parameter fits by
open circles. Note the much larger  errors for stars without 5-parameter
fits.}
\end{figure*}

\begin{figure*}
  \epsfig{file=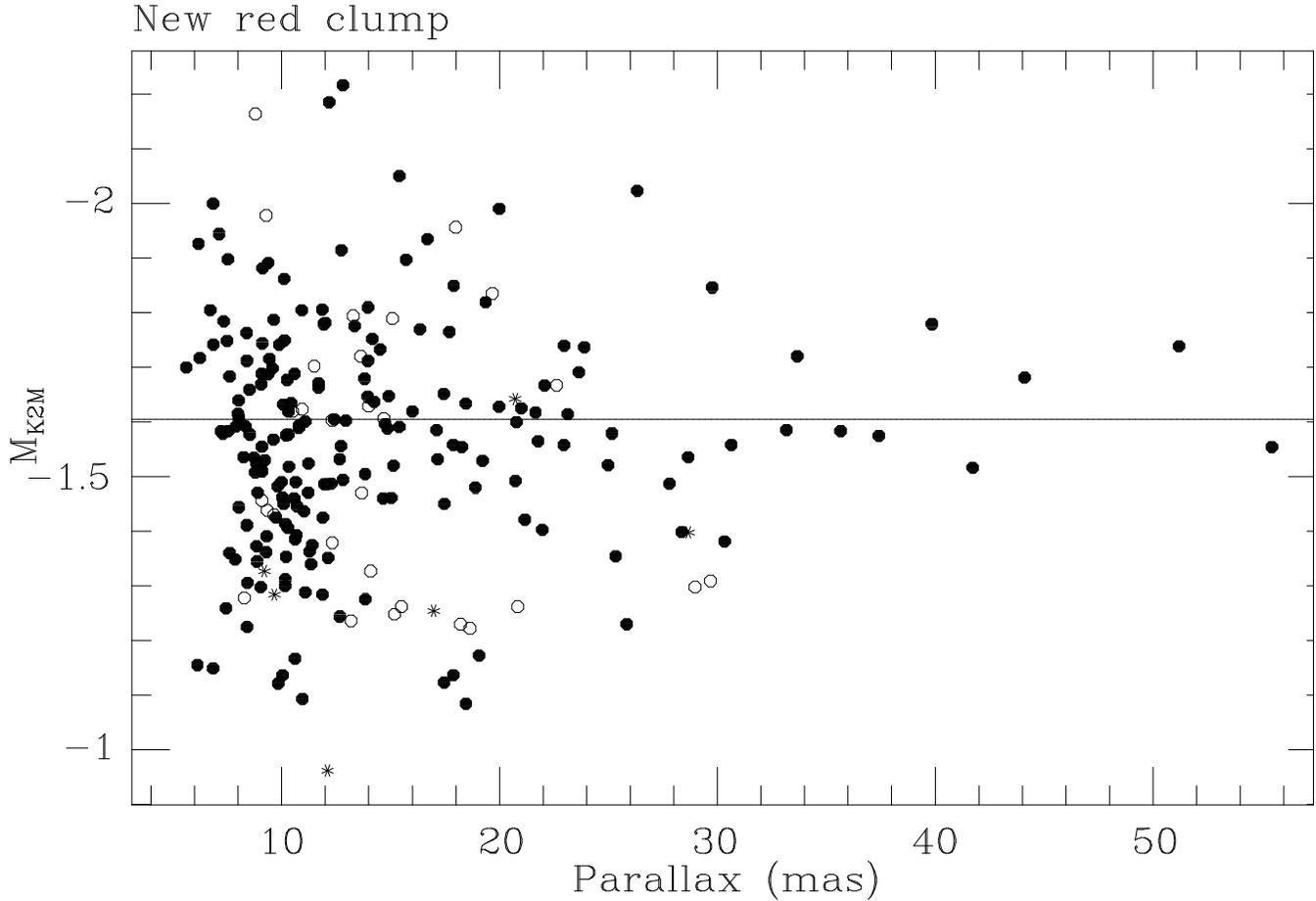, width=\linewidth}
  \caption{K$_{2MASS}$  absolute magnitude vs. parallax. Symbols as in Figure
1.The horizontal line is the mean absolute magnitude for parallaxes
$\geq$ 12 mas.}
\end{figure*}

\begin{figure*}
  \epsfig{file=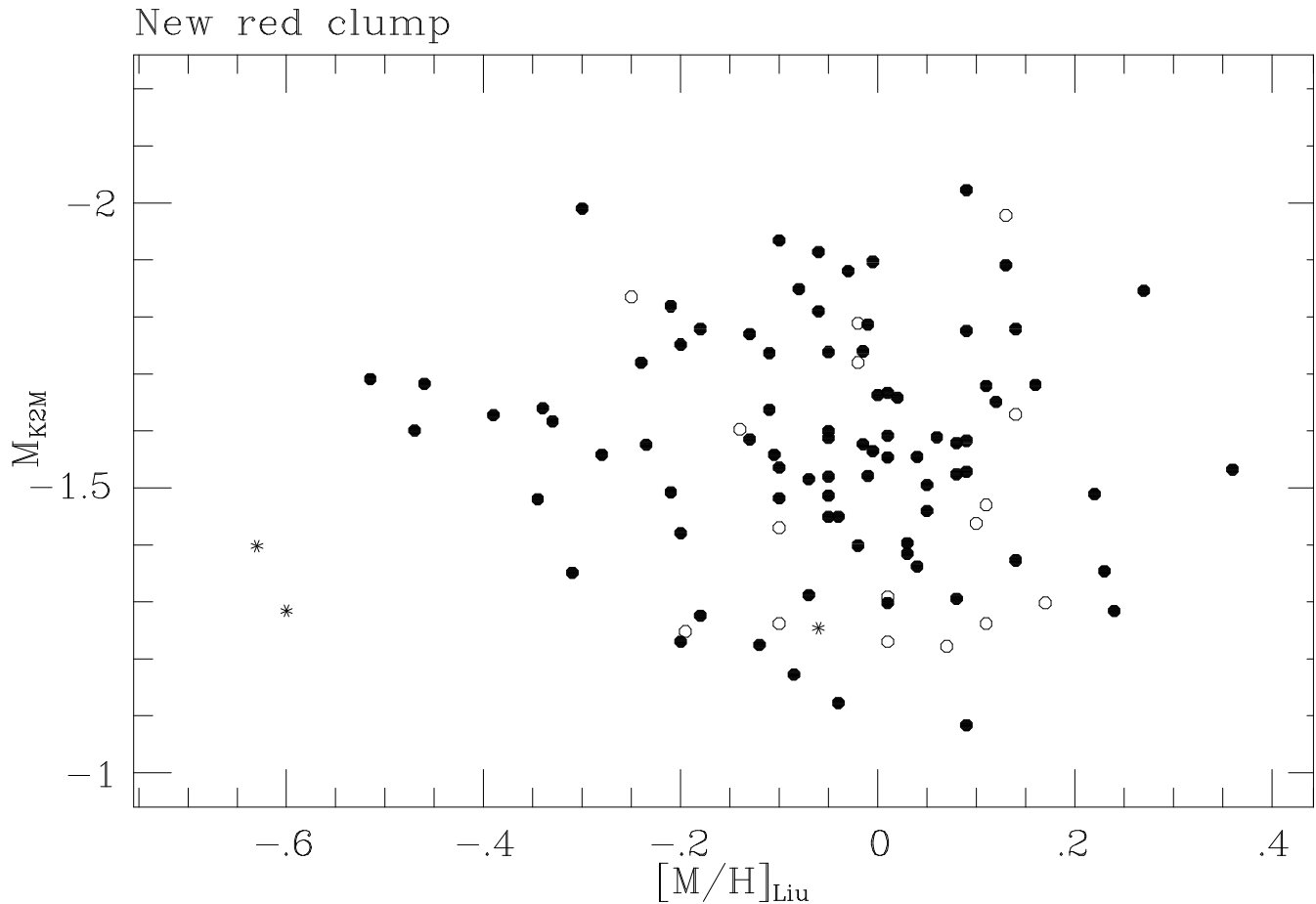, width=\linewidth}
  \caption{K$_{2MASS}$ absolute magnitudes  as a
           function of metal abundance on the scale of Liu et al. (2007).
           Stars with 5-parameter parallax fits and metal abundances are
           represented by filled circles,  and other red clump stars with
           measured metal abundances by open circles. Note the tendency
           for the stars without 5-parameter fits to be slightly fainter.}
\end{figure*}

\begin{figure*}
  \epsfig{file=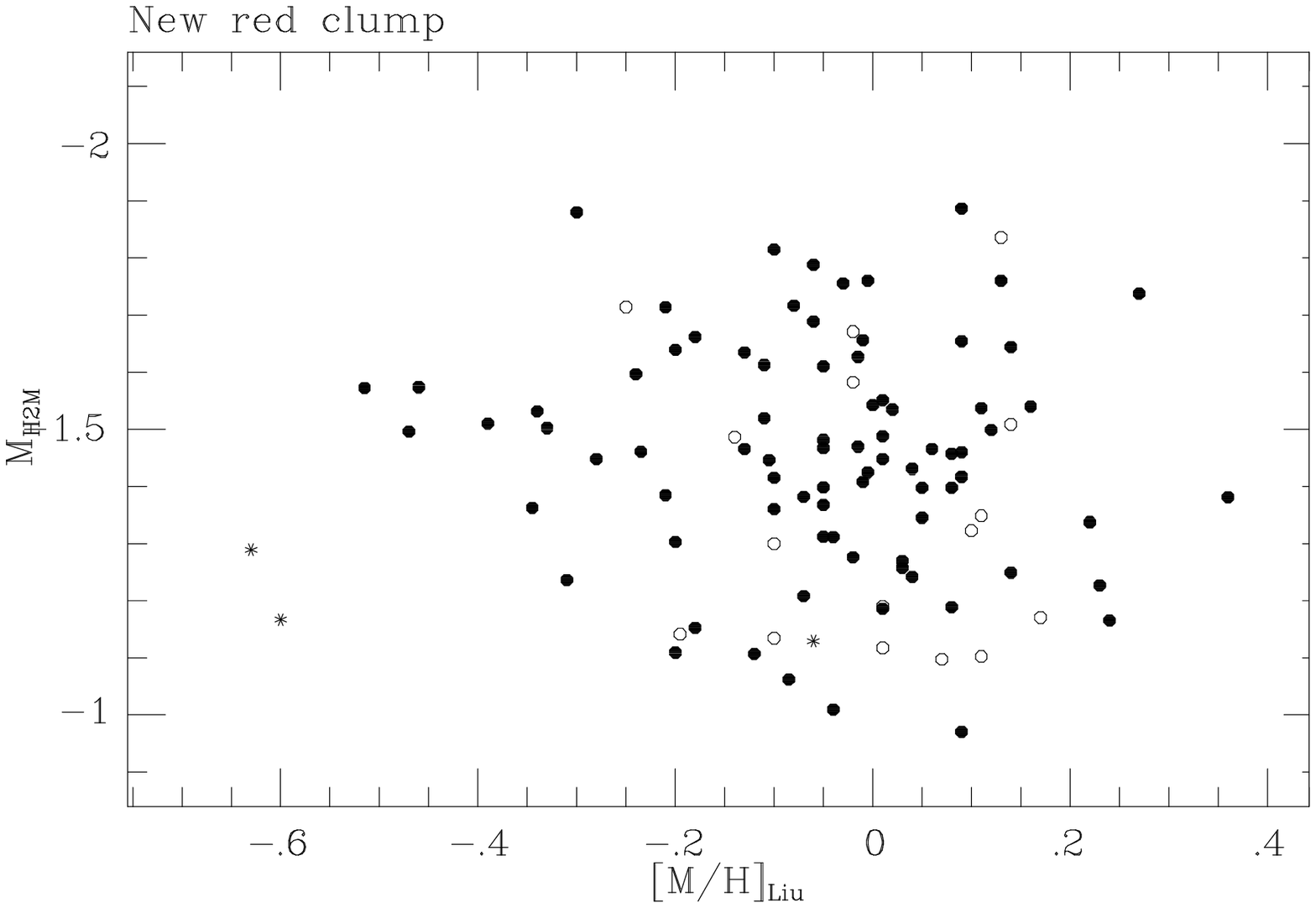, width=\linewidth}
  \caption{H$_{2MASS}$  absolute magnitudes for Hipparcos red
           clump stars as a function of metallicity. Symbols as in Figure 3.}
\end{figure*}

JHK observations for 226 nearby red clump stars with K magnitudes between
-0.3 and 4.9 were obtained with the 0.75m telescope at the South African
Astronomical Observatory (SAAO), using the Mk. II infrared photometer and
the same filter set used (Carter 1990) to define the SAAO JHKL standard
system.  Program stars were chosen from those identified by
 Paczy\'{n}ski \& Stanek (1998), selecting for declinations observable
from SAAO. As pointed out by a referee, it should perhaps be noted that 
the list of red clump stars given in Paczy\'{n}ski \& Stanek 
(1998) excluded any objects with more than 10\% error in their (original) 
Hipparcos parallaxes. As our selection from their list was not based on 
parallax, our subsample of 226 stars shares this cutoff. Our sample also 
necessarily shares their definition of the red
clump in colour and absolute magnitude.

Standard stars from the Carter list were observed frequently,
with preference given to observing standard and program stars at comparable
airmass, while minimising the angular distance between standard and program 
stars on the sky. The resulting mean JHK magnitudes (transformed to the  
2MASS system) can be
found in Table 1 (complete version online), where the transformations from the Carter (1990) system
to 2MASS have been taken from the 2MASS website (Carpenter 2003).
Of the 226 program stars, 85 were observed more than once. From this
subsample we have calculated the internal standard deviation of a single
observation to be 0.008 in J and 0.006 in H and K, while the internal mean
standard error for the 85 stars with multiple observations is 0.005 in J and
0.004 in H and K. The mean standard error for the entire sample (including
stars observed only once) is thus about 0.007 in J and 0.005 in H and K.

\begin{table*}
\begin{verbatim}
                                                      Table 1

                                 Magnitudes and absolute magnitudes on the 2MASS system
                                          for nearby Hipparcos red clump stars

                                   Apparent mag.      Hipparcos (2007)   Absolute mag.
                               ____________________  _________________  _______________
                         HIP      J      H      K     par    err  type     H        K     [M/H]**

                         671    4.282  3.757  3.653  10.16   0.42   5   -1.208   -1.312   -0.07
                        *765    2.182  1.675  1.561  22.62   0.45   1   -1.552   -1.667
                         814    3.579  3.080  2.968  12.81   0.19   5   -1.383   -1.494
                         966    4.776  4.269  4.151   8.41   0.38   5   -1.107   -1.225   -0.12
                        3137    4.150  3.612  3.485  10.62   0.43   5   -1.258   -1.385    0.03

                     *Star not used in calculating mean absolute magnitudes (see text)
                    **[M/H] on the scale of Liu et al. (2007)

\end{verbatim}
\end{table*}

The error introduced by standardisation is largely included in the above,
since the second and any additional observations of a particular red clump
star will in general not have been standardised using exactly the same
choice of standards as for the first observation of that star.

How large is the error introduced by random errors in the standards used? A
comparison of the Carter (1990) and CIT standards can be used to estimate
the error introduced by standardisation. Assuming equal errors in both
standard sets, the transformation equations given by Laney \& Stobie (1993)
imply a mean error in H and K of approximately 0.006. In general, there
will be two standards involved in standardising a given programme star,
thereby reducing the standardisation error to roughly 0.004, which suggests
that random errors and standardisation errors are of about the same
magnitude. Our precision and accuracy should be more than adequate
for the present purpose.

\section {Deriving mean absolute magnitudes for the Hipparcos sample}

The absolute magnitudes given in Table 1 for our sample of nearby red clump
stars were derived using the current Hipparcos parallaxes
(van Leeuwen 2007), assuming as did Paczy\'{n}ski \&  Stanek (1998) and Alves
(2000) that reddening is of negligible importance in the near-infrared for 
these nearby stars (average distance less than 70 pc). Examination of the
figures in Marshall et al. (2006) also suggests that extinction in K 
is likely to be negligible. As a further test, we looked for a trend with
parallax in H-K and J-K. No significant trend with parallax was found for
J-K, while the trend in H-K suggests a K-band extinction only 0.003 greater
for the most distant stars in our sample (about 180 pc) compared to the 
nearest (less than 20 pc), which is not too surprising given that almost 
all the stars in our sample lie within the 'local bubble' radius given by
Jones, West \& Foster (2011).  

Some of the 226 stars observed were not used in deriving {\it mean} 
absolute magnitudes. It can be seen from Fig. 1 that the 29 stars
for which Hipparcos parallaxes with 5-parameter fits (type 5) are not 
available (van Leeuwen 2007) tend to
have parallaxes with substantially larger error bars. These 29 stars have
therefore been omitted. Of the remaining 197 stars, six were also omitted
from our final 'good' list -- one with the reddest J-H color (and hence
possibly reddened), one with a less than optimal parallax fit in the original
Hipparcos reduction, two with low metal abundances outside the range of the
rest of the sample, one whose absolute magnitude was clearly an outlier for
its colors, and one with an abnormally large parallax error. Including these
stars would decrease the mean absolute magnitude in both H$_{2MASS}$ and
K$_{2MASS}$ by about 8 mmag.

The sample we actually used in determining mean absolute magnitudes
therefore includes 191 of the 226 stars observed. Among these stars,
there is a slight (2$\sigma$) tendency (Fig. 2) for the
stars with parallaxes lower than about 12 mas to give fainter absolute
magnitudes, which is in the expected sense for a sample with a cutoff
determined either by parallax or by percentage error in parallax.
Note that below 12 mas the parallax error begins to increase
markedly (Fig. 1). If we define f to be 1 for parallaxes less than 12 mas
and 0 for parallaxes $\geq$ this value, we can write
\begin {equation}
M_K  = -1.605\pm0.022 + 0.062\pm0.030f
\end {equation}
\begin {equation}
M_H = -1.481\pm0.022 + 0.062\pm0.029f
\end {equation}
\begin {equation}
M_J = -0.974\pm0.020 + 0.057\pm0.027f
\end {equation}
where K, H and J are on the 2MASS system.

Calculation of the effects of Lutz-Kelker correction (Smith 1999) for the
86 stars with parallaxes (van Leeuwen 2007) of 12 mas or greater gives a
very small mean correction, which raises the mean K, H and J absolute
magnitudes for this 'large parallax' subset to
-1.607$\pm$0.022, -1.484$\pm$0.022 and -0.976$\pm$0.020, respectively.

Since Lutz-Kelker bias is a selection effect, and we selected stars
with {\it revised} (i.e. 2007) Hipparcos parallaxes less than 12 mas, our
calculation of the Lutz-Kelker corrections was likewise based on the revised
Hipparcos parallaxes and errors. But our complete sample (191 stars) shares
the cutoff in the original list of Paczy\'{n}ski \& Stanek (1998), of which
our observing list was a southern subset. This cutoff was based on the
{\it original} (1997) Hipparcos results, and our calculation of the
Lutz-Kelker corrections for our sample of 191 stars must likewise be based
on the original Hipparcos parallaxes and errors. Such a calculation gives
\begin {equation}
M_K  = -1.613\pm0.015
\end {equation}
\begin {equation}
M_H = -1.490\pm0.015
\end {equation}
\begin {equation}
M_J = -0.984\pm0.014
\end {equation}
on the 2MASS system. Reassuringly, these results differ from those derived
using our 'large parallax' subset (see above) by only 6-8mmag. Likewise
reassuring is the fact that a comparison of these results with those
including only the nearer stars shows no sign whatever of extinction effects.

\section {Trends with metal abundance}

Given past interest in the effect of metal abundance on red clump absolute
magnitudes, we examined our data to see if any trend was apparent.

From our sample, 101 stars had metal abundances either from McWilliam (1990)
or Liu et al. (2007). A comparson of 24 stars in common showed that the
abundances from these two sources had different zero points, and
that the McWilliam metallicities could be placed on the scale
of Liu et al. simply by adding 0.12$\pm$0.02. For the stars with abundances
from both sources, the two values have been averaged.

Absolute magnitudes in K$_{2MASS}$ and H$_{2MASS}$ have been plotted against 
metallicity (on the
Liu et al. scale) in Figs. 3 and 4. As is evident from these figures, there
is no strong, significant trend in either absolute magnitude with 
metallicity (at least for stars with [M/H] greater than -0.6), in agreement 
with the result found by Alves (2000).

\section{LMC distance modulus}

To derive an LMC distance modulus, we need a mean K$_{2MASS}$ value for
LMC red clump stars. The two most comprehensive studies of LMC red clump
stars in the field give dereddened mean K$_{2MASS}$ magnitudes of
16.887$\pm$0.009 (Alves et al. 2002) and 16.897$\pm$0.009
(Szewczyk et al. 2008),  while an extensive recent survey of LMC red clump
stars in clusters (Grocholski et al. 2007) yields a mean of 16.891$\pm$0.032.
These are in extremely good agreement, and have been averaged to give
\begin{equation}
                      K_{2MASS} = 16.892\pm0.011
\end{equation}
We have followed Grocholski et al. (2007) in using the reddening
law of Cardelli et al. (1989). The mean K magnitudes from  Alves et al. (2002) 
and Grocholski et al. (2007) have been corrected to the LMC center as 
described in those papers, while the mean K magnitude from Szewczyk et al. 
(2008) has been left uncorrected for reasons cited by the authors of that 
paper. In all cases the K magnitudes from these three papers have been 
transformed to the 2MASS system using the transformations on the 2MASS 
website (Carpenter 2003). The mean K magnitudes for red clump stars in 
individual LMC clusters as given in  Grocholski et al. have not been 
corrected for age and metallicity according to the formulation given by 
those authors, as this actually increases the dispersion in distance modulus.

Combining the mean K$_{2MASS}$ magnitudes for the solar
neighbourhood and the LMC gives an uncorrected LMC distance modulus of
18.505$\pm$0.019. Applying K-band corrections for the age and metallicity
differences between LMC and solar-neighborhood red clump populations of
-0.03 (Salaris \& Girardi 2002) gives a 'true' LMC K-band distance modulus
of 18.475$\pm$0.021, where we have somewhat arbitrarily
allowed for an uncertainty of 0.01 mag in the population correction.

The only H-band mean magnitude for LMC red clump stars
currently available in the literature is $H_{2MASS} = 17.03\pm0.06$ 
from Koerwer (2009).
On the assumption that H-K is about equal in LMC and solar neighborhood red
clump stars (given that H-K is insensitive to both temperature and
metallicity), we apply the same population correction as for K to get an
H-band LMC modulus of 18.49$\pm$0.06, in good agreement with the K-band value
but much less tightly constrained.

For LMC red clump stars in the J band,
we have used the values given by Szewczyk et al. (2008), as
neither Alves et al. (2002) nor Grocholski et al. (2007) provide J-band
measurements. We get a mean J$_{2MASS}$ for LMC red clump stars of
17.40$\pm$0.02, and hence an {\it uncorrected} LMC distance modulus of
18.38$\pm$0.03. The discrepancy between this and the corrected K-band
modulus
is unsurprising, since the mean J-K for red clump stars in the LMC is about
0.13 bluer than in the solar neighbourhood, indicating
that a substantial population correction would be required. Our results
suggest that this correction would lie roughly halfway between the value for
I (0.2, Girardi \& Salaris 2001) and K (-0.03, Salaris \& Girardi 2002).

\section {Comparison with other results}

The mean red clump absolute magnitude in K derived above is consistent
with that of Alves (2000), although an exact comparison is
difficult given the absence of a well-defined standard system in the data
used there. Our result is, however, somewhat brighter than that derived
by Groenewegen (2008). This is not surprising in view of the trend toward
fainter absolute magnitude with decreasing parallax seen in our own data.
Alves' sample included a range of parallaxes similar to that used here,
while Groenewegen's sample included only stars more distant than those
we observed. The hypothesis considered by Groenewegen, that a bias might
be present in his result because of a lack of data for bright nearby red
clump stars, is thus confirmed. As all three studies use the same definition
of the red clump (i.e. Paczy\'{n}ski \& Stanek 1998), this is not a
factor in the comparison.

The {\it corrected} distance to the LMC derived above is in good agreement
with the K-band red clump distance derived by Alves et al. (2002)
(18.49$\pm$0.03), which includes the same Salaris \& Girardi (2002)
population correction. Our {\it uncorrected} distance is in excellent agreement
with Pietrzy\'{n}ski, Gieren and Udalski (2003) (18.50$\pm$0.01), who applied
no correction for population differences. Red clump LMC distances derived
using V and I magnitudes would
need much larger and more uncertain corrections for abundance and age
effects, and are best excluded from comparison (Pietrzy\'{n}ski et al.
2010). While the K-band correction undoubtedly has some uncertainty
attached, the correction itself is quite small.

The {\it uncorrected} K-band Cepheid distance moduli
of 18.48$\pm$0.04 (Benedict et al. 2007) and 18.47$\pm$0.03
(van Leeuwen et al. 2007) are likewise in excellent agreement, especially
with our {\it corrected} distance. The uncorrected V-band and W$_{VI}$
distance moduli from Benedict et al. (2007), 18.50 $\pm$0.03 and
18.52$\pm$0.06, and the W$_{VI}$ modulus from van Leeuwen et al. (2007),
18.52$\pm$0.03, are slightly larger, but LMC Cepheids have long been known to
be bluer at a given period than Cepheids in the solar neighbourhood (Gascoigne
\& Kron 1965, Laney \& Stobie 1986, 1994), so this small difference is in the
expected sense, though hardly significant.

Agreement with the Cepheid moduli apparently also implies good agreement
with the most recent RR Lyrae results from HST parallaxes
(Benedict \& McArthur 2011). For Type
II Cepheids, the latest results (Matsunaga, Feast \& Menzies 2009) give
18.46$\pm$0.10, which is in very good agreement although considerably less
precise. This supersedes the earlier result (Feast et al. 2008), which gave a
rather smaller modulus.

Results from LMC eclipsing binaries are rather sparse, and only one result
is available for a binary where empirical surface brightnesses are available
(Pietrzy\'{n}ski et al. 2009). Agreement between their value for the
LMC modulus (18.50$\pm$0.06) and ours is reasonable enough,
but a final comparison will have to wait until results for the remaining
seven binaries in that phase of the Araucaria Project are available.

\section {Conclusions}

Near-IR observations of 226 red clump stars as bright as $K =
-0.3$ have resulted in a determination of the local mean absolute magnitude
in H$_{2MASS}$ and K$_{2MASS}$ accurate to $\pm0.02$ mag. A comparison with K-band absolute
magnitudes for LMC red clump stars from the literature implies an LMC
distance modulus of 18.50$\pm0.02$ (uncorrected), or 18.47$\pm0.02$
(corrected by the value given in Salaris \& Girardi 2002).

Comparison of this result to {\it uncorrected} Cepheid PL-relation distance
moduli in the K-band (van Leeuwen et al. 2007, Benedict et al. 2007) suggests
that metallicity corrections to distance moduli derived from near-IR Cepheid
PL relations may not be very significant, at least for abundances between
those in the solar neighbourhood and in the LMC.

In addition, the agreement between our distance modulus
and those derived from Cepheid  W$_{VI}$ PL relations
(van Leeuwen et al. 2007, Benedict et al. 2007) suggests that Bono et
al. (2010) may be correct in arguing that metallicity corrections to
distances from Cepheid Wesenheit (VI) PL relations may be fairly negligible.

Much the same holds for the V-band PL relation (Benedict et al. 2007), and
these conclusions are strengthened by the recent results from RR Lyraes,
Type II Cepheids, and (on a preliminary basis) one late-type double-line
eclipsing binary as mentioned above.

In turn the results found here and in other recent papers strengthen the
conclusions reached by Bresolin (2011), who found such a shallow oxygen
abundance gradient in NGC 4258 HII regions that abundance differences could
not realistically explain the difference in brightness between Cepheids in
the outer and inner regions of that galaxy.

\section{Acknowledgements}

We would like to thank the South African Astronomical Observatory for
generous allotments of observing time with what may be the world's last
instrument capable of high-precision IR observations of bright objects.  
Travel funding for this project has been provided by the Brigham Young
University Department of Physics and Astronomy.  Support from the FOCUS 
and TEAM subsidies of the Foundation for Polish Science (FNP) is also 
acknowledged.  We extend our thanks to Dr. Benjamin J. Taylor for his
comments on our original manuscript.  We also thank Lisa Joner for her 
careful proofreading of several different versions of this paper. This
research has made use of the VizieR catalogue access tool, CDS, Strasbourg,
France.

\end{document}